\documentclass[10 pt,conference, letterpaper]{IEEEtran}
\IEEEoverridecommandlockouts
\usepackage{times,amsmath,epsfig,latexsym,amssymb,psfrag,eufrak,dsfont}

\usepackage{cite}
\usepackage{booktabs}
\usepackage{paralist}
\usepackage[ruled,vlined]{algorithm2e}
\usepackage{float}
\restylefloat{table}
\usepackage{graphicx}
\usepackage{caption}
\usepackage[font=small]{caption}
\usepackage{subcaption}
\usepackage{float}
\floatstyle{plaintop}
\restylefloat{table}

\usepackage{color}

\usepackage{subcaption}
\usepackage[utf8]{inputenc}
\newcommand{\qed}{\nobreak \ifvmode \relax \else
\ifdim\lastskip<1.5em \hskip-\lastskip
\hskip1.5em plus0em minus0.5em \fi \nobreak
\vrule height0.75em width0.5em depth0.25em\fi}
\newtheorem{rem}{Remark}

\usepackage{upquote}

\def\({\left(}
\def\){\right)}

\def\[{\left[}
\def\]{\right]}

\begin{document}
\graphicspath{{figures/}}
\title{Cooperative Transmissions in Ultra-Dense Networks under a Bounded Dual-Slope Path Loss Model}

%\author{\IEEEauthorblockN{Yanpeng Yang\IEEEauthorrefmark{1}, Jihong Park\IEEEauthorrefmark{2}, Seong-Lyun Kim\IEEEauthorrefmark{3} and Ki Won Sung\IEEEauthorrefmark{1}}
%\IEEEauthorblockA{\IEEEauthorrefmark{1}KTH Royal Institute of Technology, Wireless@KTH, Stockholm, Sweden\\
% E-mail: yanpeng@kth.se, sungkw@kth.se}
%\IEEEauthorblockA{\IEEEauthorrefmark{2}Department of Electronic Systems, Aalborg University, Denmark\\
%Email: jihong@es.aau.dk}
%\IEEEauthorblockA{\IEEEauthorrefmark{3}Dept. of Electrical \& Electronic Enginerring, Yonsei University, Seoul, Korea\\
%Email: slkim@ramo.yonsei.ac.kr}
%}

\author{\IEEEauthorblockN{Yanpeng Yang and Ki Won Sung}
\IEEEauthorblockA{Wireless@KTH\\
KTH Royal Institute of Technology, Sweden\\
Email: $\{$yanpeng, sungkw$\}$@kth.se}
\and
\IEEEauthorblockN{Jihong Park}
\IEEEauthorblockA{Dept. of Electronic Systems\\
Aalborg University, Denmark\\
Email: jihong@es.aau.dk}
\and
\IEEEauthorblockN{Seong-Lyun Kim and Kwang Soon Kim\\}
\IEEEauthorblockA{School of Electrical \& Electronic Enginerring\\
Yonsei University, Seoul, Korea\\
Email: \{slkim, ks.kim\}@yonsei.ac.kr}}

\IEEEpubid{978--1--5386--3873--6/17/\$31.00~\copyright~2017 IEEE }
\maketitle

\begin{abstract}
%\boldmath
In an ultra-dense network (UDN) where there are more base stations (BSs) than active users, it is possible that many BSs are instantaneously left idle. Thus, how to utilize these dormant BSs by means of cooperative transmission is an interesting question. In this paper, we investigate the performance of a UDN with two types of cooperation schemes: non-coherent joint transmission (JT) without channel state information (CSI) and coherent JT with full CSI knowledge. We consider a bounded dual-slope path loss model to describe UDN environments where a user has several BSs in the near-field and the rest in the far-field. Numerical results show that non-coherent JT cannot improve the user spectral efficiency (SE) due to the simultaneous increment in signal and interference powers. For coherent JT, the achievable SE gain depends on the range of near-field, the relative densities of BSs and users, and the CSI accuracy. Finally, we assess the energy efficiency (EE) of cooperation in UDN. Despite costing extra energy consumption, cooperation can still improve EE under certain conditions.

%Ultra-dense network (UDN) is believed to be one of the essential elements of 5G wireless systems. 
\end{abstract}



% \fi
\IEEEpeerreviewmaketitle

\section{Introduction}

Mobile communication technologies are rapidly prompted by the tremendous growth of traffic demand. Deploying massive number of cheap small base stations (BSs), so called an ultra-dense network (UDN), represents a paradigm shift from conventional deployment strategies\cite{Kamel1}\cite{Lopez}. Compared with the traditional networks designed for fully loaded operation, UDN is partially loaded in its inherent design because the BS density exceeds that of users \cite{Lopez}\cite{Park}.

%\begin{figure}[t]
%\includegraphics[width=.5\textwidth]{Intro2.pdf}
%\caption{BS cooperation in (a) Traditional fully loaded network, (b)Partially loaded UDN.}
%\label{fig:Intro}
%\end{figure}

%\begin{figure}[t!]
%\centering
%\begin{subfigure}
%{\label{fig:TCoMP}
%\includegraphics[width=0.5\textwidth]{Intro1.pdf}}
%\end{subfigure}
%\begin{subfigure}
%{\label{fig:UCoMP}
%\includegraphics[width=0.5\textwidth]{Intro2.pdf}}
%\end{subfigure}
%\caption{Cooperative transmission in (a) Traditional fully loaded network, (b)Partially loaded UDN.}
%\label{fig:CoMP}
%\end{figure}

\begin{figure}[ht]
  \centering
  \begin{subfigure}[a]{0.5\textwidth}
    \centering\includegraphics[width=0.9\textwidth]{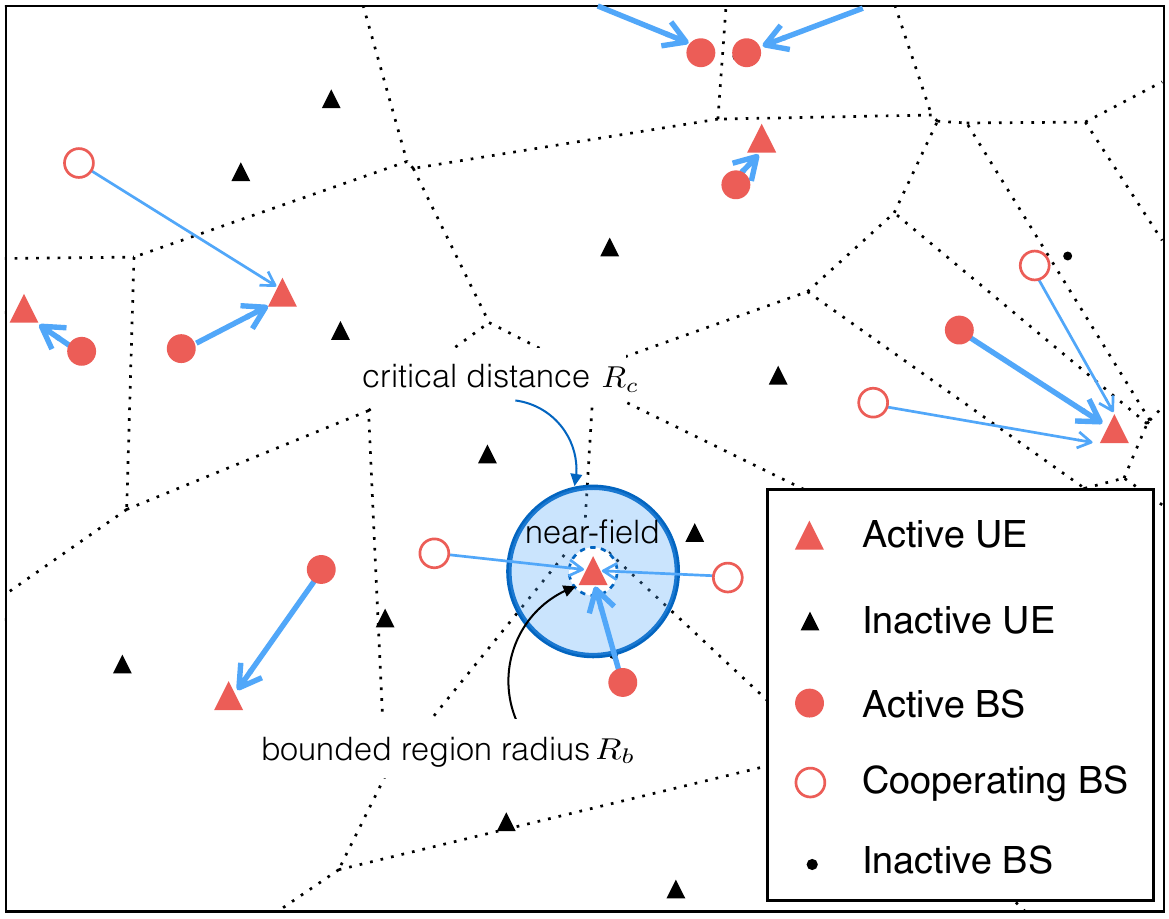}
    \caption{Traditional joint transmission}
    \label{fig:TCoMP}
  \end{subfigure}\\
  \begin{subfigure}[a]{0.5\textwidth}
    \centering\includegraphics[width=0.9\textwidth]{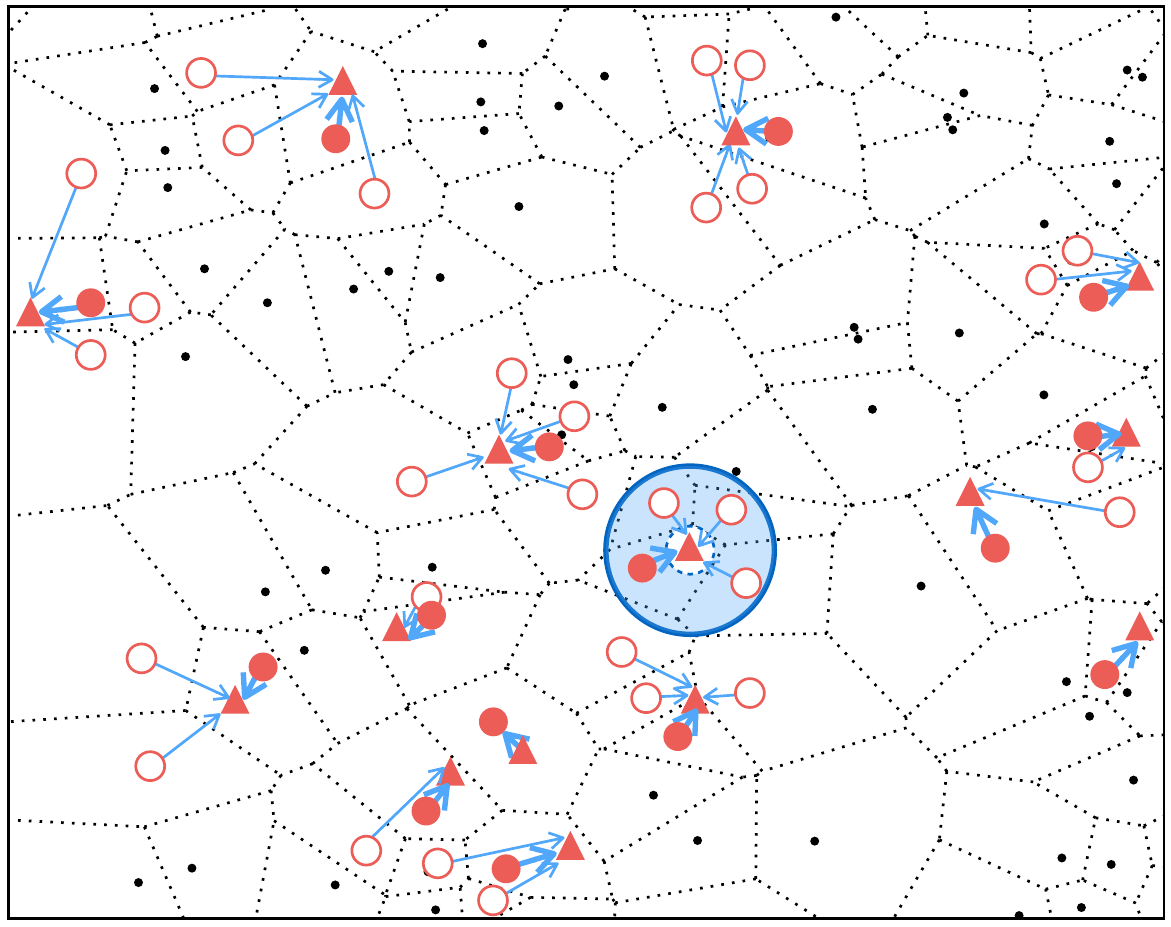}
    \caption{UDN joint transmission}
    \label{fig:UCoMP}
  \end{subfigure}
  \caption{An illustration of joint transmission in (a) traditional fully-loaded network and (b) partially-loaded UDN.}
   \label{fig:CoMP}
\end{figure}

\subsection{Motivation and Related Work}
The existing studies of UDN mainly focus on the single BS association \cite{Park, Yang2}. The BSs without user in their coverage areas are considered in sleep mode to save energy and reduce interference. In extreme cases, a large number of BSs will temporarily stay idle in the network. Thus, it raises a research question if we can exploit such temporary infrastructure 'redundancies' in UDN to improve the system performance.

Joint transmission (JT) is a potential solution which allows multiple BSs to jointly serve one user. In traditional fully loaded cellular networks, JT could turn dominant interferers into useful signals as shown in Fig. \ref{fig:CoMP}(a) while the other interferers remain the same. Thus, the desired signal strength increases and the interference decreases simultaneously, at the cost of reduced scheduling probability. 
\IEEEpubidadjcol
It is known that JT enhances the performance of cell-edge users in macro cellular networks \cite{Marsch}. However, interference nature is completely different in a UDN because turning on dormant BSs is like a double-edged sword, i.e., improving the desired received signal strength, but generating extra interference and energy consumption. In Fig. \ref{fig:CoMP}(b), if all the users get assistance from nearby sleeping BSs, the interference will grow rapidly as well as the desired signal power. Therefore, how to design cooperation schemes in UDN to overcome the concurrent interference becomes a big challenge. A cooperative UDN architecture is proposed in \cite{Chen}, but without further discussions on cooperation schemes and performance evaluation.

To examine the impact of JT on UDN, it is important to incorporate the propagation characteristics of UDN properly. In a UDN environment where the cell sizes are getting much smaller, a widely accepted unbounded single-slope path loss model, i.e., $G(d) = d^{-\alpha}$, becomes dubious. The radio signals in the near-field may experience much less absorption and diffraction losses than those in the far-field, resulting in dissimilar path loss exponents. Besides, the probability of a link within a reference distance, $\emph{d} \in (0,1)$, becomes high, and thus this phenomenon cannot be neglected in the analysis. Hence, a path loss model with multiple slopes and bound becomes necessary in modeling the UDN scenario. The impact of bounded and multi-slope path loss models in fully loaded networks are separately studied in \cite{Inaltekin} and \cite{Zhang}\cite{Ding}. However, the combination of the two effects remains to be explored. Moreover, the full load assumption becomes implausible in the UDN environment since the BS density exceeds the user density \cite{Yu}\cite{Yang}.

\subsection{Contributions}
This paper intends to give a first look at applying BS cooperation in a partially loaded UDN scenario. We employ a bounded dual-slope path loss model in order to capture the characteristics of UDN. Furthermore, two cooperation schemes are investigated: non-coherent JT without the assistance of instantaneous channel state information (CSI) and coherent JT with full CSI knowledge. Our key findings on UDN cooperation are summarized as follows:
\begin{itemize}
\item Exploiting CSI is necessary for cooperative transmissions in UDN (Remark 1).
\item Cooperation gain in spectral efficiency (SE) increases with the range of near-field, i.e critical distance (Remark 2), as well as both near/far-field path loss exponents (Remark 3).
\item Cooperation gain also grows with active user density (Remark 4 and Fig. \ref{fig:UEdensity}), but is convex-shaped over BS density (Remark 5).
\item With imperfect CSI, cooperation is more preferable under lower operating frequency.
\item Cooperation can also increase network energy efficiency (EE) within a limited number of cooperating BSs.
\end{itemize}

%The remainder of this paper is organized as follows: The system model is explained in Section \ref{sec:SM}. Then, we introduce our cooperation schemes in Section \ref{sec:Scheme}. In Section \ref{sec:Cooperation}, we present the numerical results and analysis for cooperation in a UDN. Finally, the conclusions are discussed in Section \ref{sec:Con}.

%%%%%%%%%%%%%%%%%%%%%%%%%%%%%%%%%%%%%%%%%
\section{Network and Channel Models}
\label{sec:SM}
\subsection{Network Setup}

\begin{figure}[t]
\includegraphics[width=.45\textwidth]{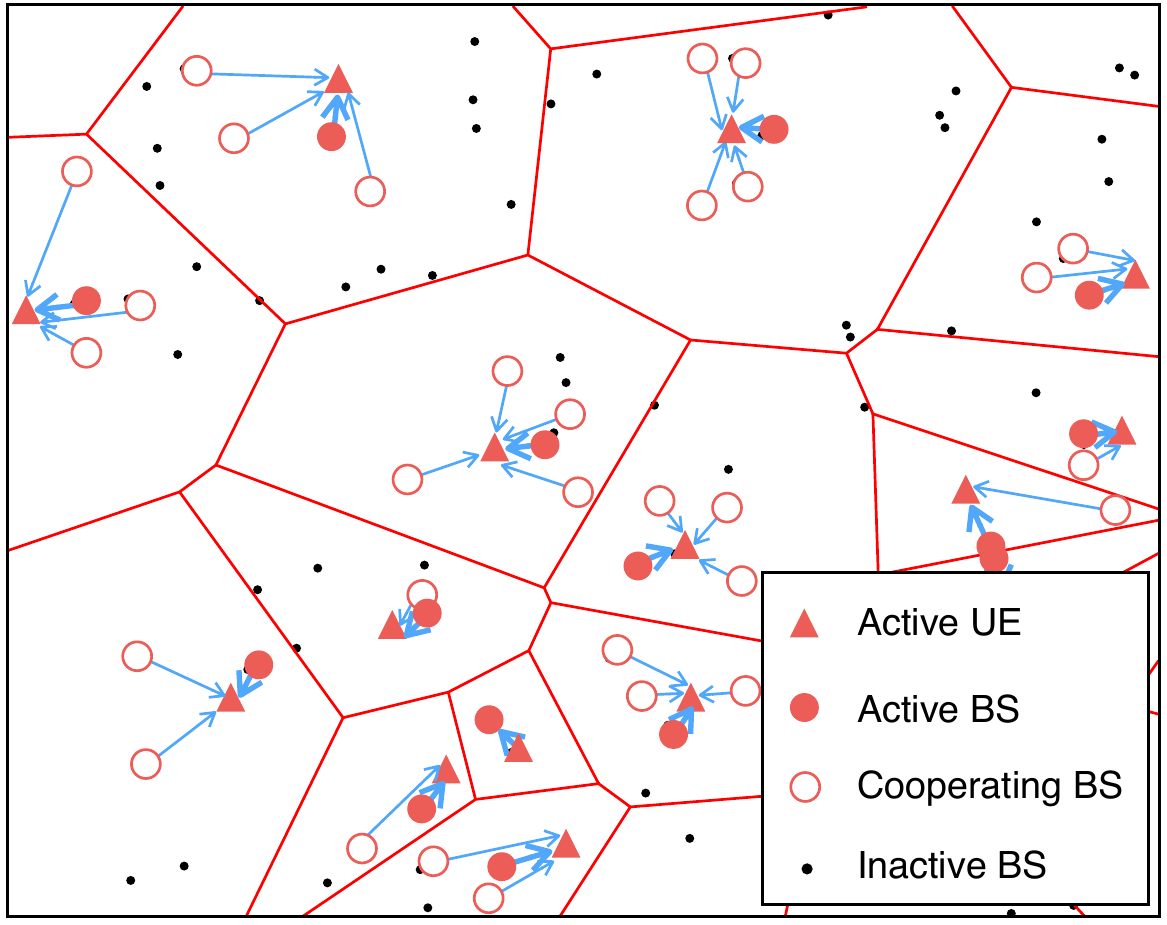}
\centering
\caption{User-centric Voronoi cells for a modeling of cooperative transmissions in a UDN.}
\label{fig:Voronoi}
\end{figure}

We consider the downlink of a UDN with BS density $\lambda_{b}$ and user density $\lambda_{u}$ that follow independent Poisson Point Processes (PPPs) $\Phi_{b}$ and $\Phi_{u}$, respectively. According to the definition of UDN, we set $\lambda_{b} \gg \lambda_{u}$ as in \cite{Park}\cite{Yu}. This can be interpreted as off-peak hours under extremely dense BS deployment. It is also supported by real traffic measurement \cite{MeanLoad} where only up to 20\% BSs are active to make traffic queues stable.

Both BSs and users are equipped with single antenna and BS transmits with unit power. Each user is associated with its closest BS when no cooperation occurs. Each BS becomes dormant when its coverage is empty of users. Dormant BSs do not transmit signals, i.e. not interfering with others, but consume energy to be specified in Section II-C. We assume Rayleigh fading in this work with the fading coefficients $h$ to be i.i.d. complex normal distributed random variables with zero mean and unit variance. 
%Shadow fading is ignored for brevity since it will not affect the major conclusions.

In previous studies \cite{Zhang}\cite{Yu}, the network topology is composed by BS-centric Voronoi cells which represent the BS coverage areas as shown in Fig. \ref{fig:CoMP}(a). However, for the case of UDN, massive empty cells result in an ineffective map partition. Concerning this, we propose to reverse the roles of BS and user in the topology and define user-centric Voronoi cells for modeling the cooperation in UDN, as shown in Fig. \ref{fig:Voronoi}. In cooperation model, up to $N$ BSs in the Voronoi cell will jointly serve one user. Since the system can be approximated as interference-limited in dense networks \cite{Park2}, we will neglect the noise power and examine signal-to-interference-ratio (SIR) throughout the paper.

%The BSs inside a user-centric Voronoi cell is closer to the particular user than to other users. 

%Moreover, based on the UDN density assumption, we can approximate the $k$th closest BS inside the Voronoi cell as the $k$th closest BS in the entire network when $k \ll  \frac{\lambda_b}{\lambda_u}$.

\subsection{Path Loss Model}

We apply a bounded dual slope path loss model in this work. The model divides the entire region into three parts: bounded region, near-field area, and far-field area (Fig. \ref{fig:CoMP}(a)). Bounded region is a closed circle centered at the user, inside which the path loss is assumed constant. It is to avoid received power larger than transmitted power in a short distance. Outside the bounded region, the signal experiences different path loss exponents in near-field and far-field areas, divided by the critical distance. The model can be expressed as in \cite{Zhang, Yang2}:
\begin{equation}
\label{equ:pathloss}
\ell(\alpha_{1}, \alpha_{2}, x)=\left\{
        \begin{array}{ll}
          1, & 0 \leq \lVert x \rVert \leq R_{b}; \\
          \lVert x\rVert^{-\alpha_{1}}, &  R_{b} < \lVert x \rVert \leq R_{c}; \\
          \tau \lVert x\rVert^{-\alpha_{2}}, &  \lVert x\rVert > R_{c}
        \end{array}
      \right.
\end{equation}
where we set $R_{b} = 1$ as the radius of bounded path loss region for simplicity,; $\tau \triangleq R_{c}^{\alpha_{2}-\alpha_{1}}$ to keep the continuity of the function; $R_{c}  \geq R_{b}$ denotes the critical distance; $\alpha_{1}$ and $\alpha_{2}$ are the near- and far-field path loss exponents respectively, assuming $2 \leq \alpha_{1} \leq \alpha_{2}$. 

 %i.e., the path loss in the range of $[0,R_b]$ is assumed constant
 
\subsection{Power Consumption Model}
\label{sec:Power}
We assume transmitting BSs and dormant BSs are in an active-mode and a sleeping-mode. The power consumptions for active- and sleeping-mode are $P_t$ and $P_s$, respectively. We define $\theta < 1$ as the ratio between two power consumptions, i.e., $\theta = \frac{P_s}{P_t}$. By applying $N$ cooperating BSs as proposed in Section  \ref{sec:Coop}, the densities of active BSs and sleeping BSs in a unit area are $N \lambda_u$ and $\lambda_b - N \lambda_u$. Thus, the area average power consumption can be expressed as:
\begin{equation}
P_A = P_t(N\lambda_u+\theta (\lambda_b - N \lambda_u)).
\end{equation}

%%%%%%%%%%%%%%%%%%%%%%%%%%%%%%%%%
\section{Cooperative Transmission Models}
\label{sec:Scheme}

\subsection{Cooperation Scheme}
\label{sec:Coop}
Any user $i$ in the network is jointly served by the set of $N$ closest BSs in its own Voronoi cell, denoted by $\mathcal{C}_i = \bigcup_{j=1}^N \mathit{BS}_j$. Within the cooperation set, all BSs jointly transmit the same message to the user using the same frequency band. We denote $\Phi_{\mathcal{C}}$ as the set of active BSs in the whole network. Thus, the signal received by user $i$ is:

\begin{equation}
\label{eqn:signal}
y_i = \displaystyle\sum_{x\in\mathcal{C}_i}\ell(d_{x,i})^{\frac{1}{2}}h_{x,i}w_{x,i}X_i + \displaystyle\sum_{x\in\Phi_{\mathcal{C}}\backslash\mathcal{C}_i}\ell(d_{x,i})^{\frac{1}{2}}h_{x,i}w_{x,i}X_{x}
\end{equation}
where $d_{x,i}$ and $h_{x,i}$ denote the distance and channel between BS $x$ and user $i$, $h_{x,i} \sim \mathcal{CN}(0,1)$; $w_{x,i}$ is the precoder applied by BS $x$. $X_i$ and $X_x$ are the transmitted symbols sent by cooperating BSs and interfering BSs respectively.

In non-coherent JT, the receivers apply open-loop joint processing CoMP scheme as in \cite{Garcia}, where signals from different transmitters are added by power summation. The precoder $w_x$ is set to be 1 for non-coherent JT. The desired signal power for non-coherent JT is given by
\begin{equation}
\mathcal{S}_i^{\mathit{NJ}}=\sum_{x\in\mathcal{C}_i}|h_{x,i}|^2\ell(d_{x,i}).
\end{equation}

In coherent JT, we assume the CSI is available at the BS side. In this case, BSs can design precoder to adjust the phase shift of the channel and amplify the corresponding channel gain. We employ maximal ratio transmission (MRT) precoder such that $w_x=\frac{h_x^*}{|h_x|}$. The desired signal power for coherent JT is thus
\begin{equation}
\label{equ:CJpower}
\mathcal{S}_i^{\mathit{CJ}}=\left|\sum_{x\in\mathcal{C}_i}h_{x,i}w_{x,i}\sqrt{\ell(d_{x,i})}\right|^2.
\end{equation}

Therefore, the SIR of user \emph{i} with two cooperation schemes, $\gamma_i^{\!_{\mathit{NJ}}}$ and $\gamma_i^{\!_{\mathit{CJ}}}$ are shown as follows:
\begin{equation}
\label{eqn:NonJTsinr}
\gamma_i^{\!_{\mathit{NJ}}} = \frac{\mathcal{S}_i^{\mathit{NJ}}}{\displaystyle\sum_{x\in\Phi_\mathcal{C}\backslash\mathcal{C}_i}|h_{x,i}|^2\ell(d_{x,i}) }
\end{equation}

\begin{equation}
\label{eqn:CoJTsinr}
\gamma_i^{\!_{\mathit{CJ}}} = \frac{\mathcal{S}_i^{\mathit{CJ}}}{\displaystyle\sum_{x\in\Phi_\mathcal{C}\backslash\mathcal{C}_i}|h_{x,i}|^2\ell(d_{x,i}). }
\end{equation}

\subsection{Imperfect CSI}
We focus on the delayed CSI feedback caused by the movements of users. The standard Gaussian Markov process (GMP) is used to model the temporal variation of the channel state. We assume a block fading model \cite{Biglieri} where $\textbf{h}$ remains constant over a time separation and evolves thereafter according to an ergodic stationary autoregressive (AR) GMP of order 1. The channel evolves in time as:
\begin{equation}
\textbf{h}[t]=\rho \textbf{h}[t-T_s] + \textbf{e}[t]
\end{equation}
where $\textbf{h}[t]$ denotes the channel realization at time \emph{t}, $0<\rho<1$ is the channel correlation coefficient, and $\textbf{e}[t] \sim \mathcal{CN}(0,1-\rho^2)$ represents the error vector. For the coefficient $\rho$, we use Clarke's model \cite{Clarke} and set $\rho=J_{0}(2\pi f_{d} T_{s})$, where $J_{0}(\cdot)$ is the zero-th order Bessel function of the first kind, $f_{d}$ is the Doppler frequency shift, and $T_{s}$ is the time separation. Since $f_d=\frac{f_c v}{c}$, the accuracy of the feedback will highly depends on operating frequency $f_c$ and the moving speed $v$ of the user.

\subsection{Cooperation Gains}
In this study, the user SE $\mathcal{R}$ and the average network EE $\eta$ are chosen as the performance metrics. The SE derived by the Shannon formula is given by
\begin{equation}
\mathcal{R}=\log_2(1+\gamma).
\end{equation}
We set the user SE under the single association $\mathcal{R}^o$ as our baseline to measure the gain of cooperation. The SE gain $G_c$ is defined as the ratio between SE difference $\Delta_\mathcal{R}$ and $\mathcal{R}^o$ as follows:
\begin{equation}
\label{equ:Gc}
G_c=\frac{\Delta_{\mathcal{R}}}{\mathcal{R}^o}=\frac{\mathcal{R}^J - \mathcal{R}^o}{\mathcal{R}^o}
\end{equation}
where $\mathcal{R}^J$ and $\mathcal{R}^o$ are the SE with and without cooperation, respectively.

The average network EE $\eta$ is defined as the area SE divided by the average power consumption. Considering the power consumption model described in Section \ref{sec:Power}, we can express the average network EE as:
\begin{equation}
\eta = \frac{\lambda_u \mathcal{R}}{P_A}=\frac{\lambda_u \mathcal{R}}{P_t(N\lambda_u+\theta (\lambda_b - N \lambda_u))}.
\end{equation}
At last, EE gain is the ratio between the average network EE with cooperation $\eta^J$ and that EE without cooperation $\eta^o$, given by
\begin{equation}
G_\eta = \frac{\eta^J}{\eta^o}.
\end{equation}

%%%%%%%%%%%%%%%%%%%%%%%%%%%%%%%%%%%%%%%%%%%%%%%%%%%%%%%%%%%%

\section{Numerical Results and Design Guidelines for Cooperative Transmission}
\label{sec:Cooperation}
In this section, we present results for cooperation in UDN by Monte Carlo simulations. Since the potential cooperation region is the Voronoi cell of a certain user, each BS can serve at most one user. If the number of BSs in the Voronoi cell $M$ is less than the cooperation number $N$, we pick min($M$,$N$) as the cooperation number. In the following, we investigate the impacts of near/far-field channel characteristics and BS/user density on UDN cooperation, followed by the the effect of imperfect CSI and EE behaviors.

\begin{figure}[t]
\includegraphics[width=.5\textwidth]{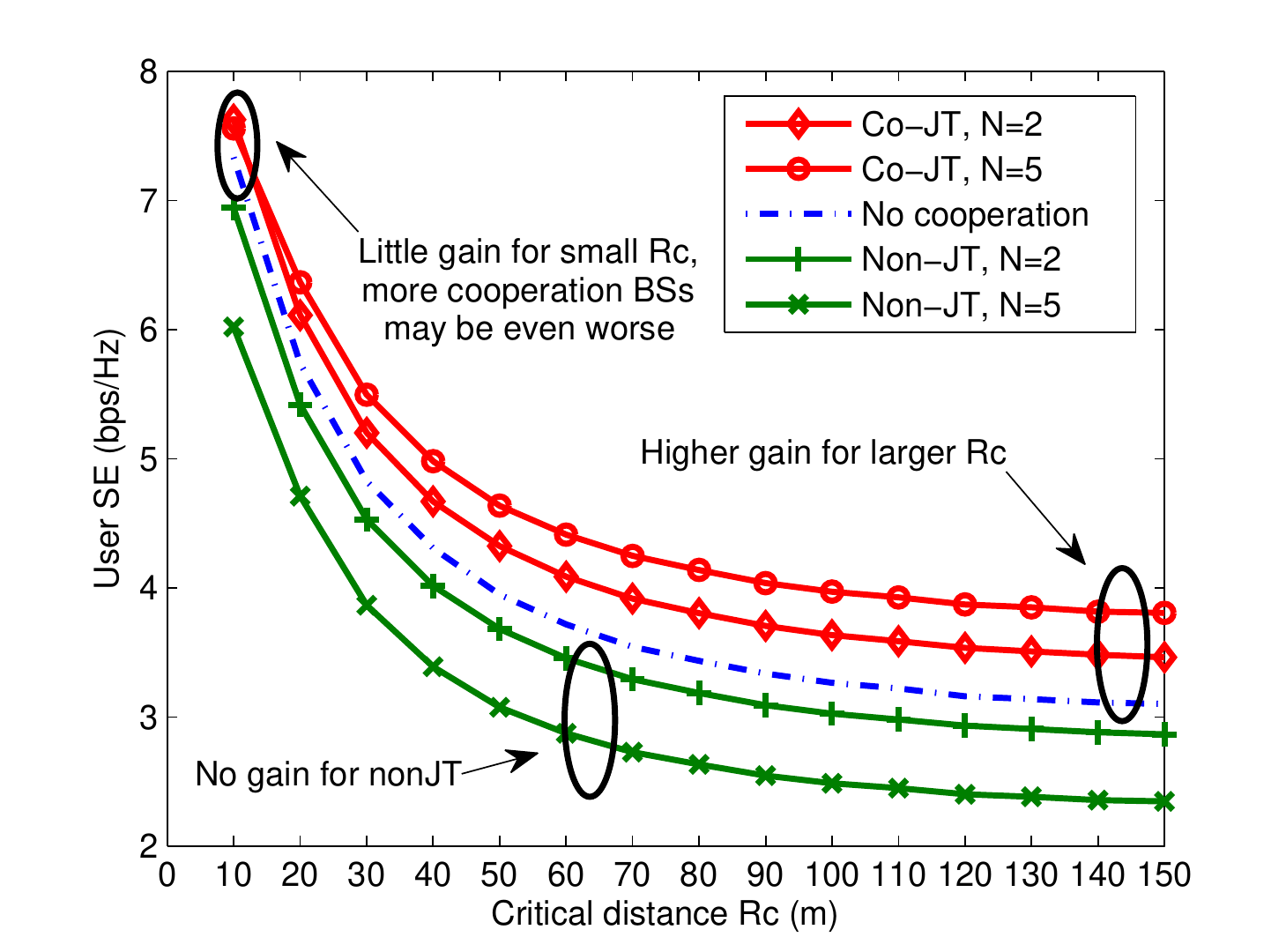}
\caption{User performance with different critical distances when $[\alpha_1,\alpha_2]$ = [2,4], $\lambda_b=4000/\text{km}^2$, $\lambda_u=200/\text{km}^2$.}
\label{fig:Rc}
\end{figure}

\subsection{Critical Distance}
\label{sec:Rc}
The effect of critical distance is illustrated in Fig. \ref{fig:Rc}. As the critical distance gets longer, all the performances decline due to more interference generated from the near-field.

\begin{rem}[Non-coherent JT]
Non-coherent JT has worse performance than single BS association.
\end{rem}

In a UDN scenario, the interferer coordinates can be approximated as the corresponding user coordinates \cite{Park}. Therefore, interference can be considered as almost linearly increasing with the number of cooperating BSs due to longer interfering distance. In non-coherent JT, the desired power summation grows diminishingly as the cooperation number increases because the transmitting distance gets longer. As a result, the increment in the desired signal is less than interference which leads to no gain. Therefore, we will not consider non-coherent JT in further discussions. All the rest of remarks are regarding coherent JT.

\begin{rem}[Effect of critical distance]
The cooperation gain grows with increasing critical distance. For a short critical distance, more cooperating BSs make SE even worse. For a large critical distance, the cooperation gain is higher with more BSs.
\end{rem}

To show the cooperation gain, we can make the following approximations:
\begin{alignat}{3}
\label{equ:analysis}
\Delta_{\mathcal{R}}&=\log_{2}(1+\gamma^{\mathit{J}})-\log_{2}(1+\gamma^o)\\
&\stackrel{(a)}{\approx} \log_{2}\(\frac{\gamma^{\mathit{J}}}{\gamma^o}\)\\
&\stackrel{(b)}{\approx} \log_{2}\(\frac{\mathcal{S}^{\mathit{J}}}{N\mathcal{S}^o}\)
\label{equ:approxb}
\end{alignat}
where (a) is because $\mathit{SIR} \gg 1$ in UDN \cite{Park3} and (b) follows from the linear relation approximation of the interference.%, $\gamma^{\mathit{J}}$ represents cooperation SIR.

When $R_c$ is small, using (\ref{equ:pathloss}) and (\ref{equ:CJpower}) in (\ref{equ:approxb}) we can get\footnote{The bounded region will not affect the conclusion. It is not considered in (\ref{equ:Rc1}) because BS density is not large enough.}:
\begin{equation}
\begin{split}
\begin{aligned}
\label{equ:Rc1}
\Delta_{\mathcal{R}}&= \log_{2}\( \frac{|\sum_{j=1}^{K}|h_{j}||d_{j}|^{-\frac{\alpha_1}{2}}+\sum_{j=K}^{N}|h_{j}|\tau^\frac{1}{2}|d_{j}|^{-\frac{\alpha_2}{2}}|^2}{N|h_{1}|^{2}|d_{1}|^{-\alpha_1}}\)
\end{aligned}
\end{split}
\end{equation}
where $K$ cooperating BSs are inside near-field and $N-K$ fall into the far-field area. The $N-K$ 'far-field' BSs can hardly make a contribution with a larger path loss exponent. Thus, incorporating more BSs into JT even decreases the gain.

%This can make the numerator smaller than the denominator which leads to $\frac{\mathcal{S}^{\mathit{J}}}{N\mathcal{S}} < 1$, thus $\Delta_{\mathcal{R}} < 0$.

When $R_c$ is large, all the cooperating BSs will fall into the near-field, then (\ref{equ:analysis}) can be written as:

\begin{equation}
\begin{split}
\begin{aligned}
\label{equ:Rc2}
\Delta_{\mathcal{R}}&= \log_{2}\( \frac{|\sum_{j=1}^{N}|h_{j}||d_{j}|^{-\frac{\alpha_1}{2}}|^2}{N|h_{1}|^{2}|d_{1}|^{-\alpha_1}}\)
\end{aligned}
\end{split}
\end{equation}
which is independent of $R_c$ because both the numerator and denominator are inside near-field. However, a larger $R_c$ leads to a lower baseline $\mathit{R}^o$ which results in a higher cooperation gain. In our simulation, a $19.6\%$ gain is obtained by 5 coherent cooperating BSs with $R_c=70$m while it increases to $23\%$ when $R_c=150$m.

\subsection{Path Loss Exponent}
\label{sec:PLexponent}

\begin{rem}[Effect of path loss exponent]
The cooperation gain decreases with both near-field and far-field path loss exponents.
\end{rem}

\begin{figure}[t]
\includegraphics[width=.5\textwidth]{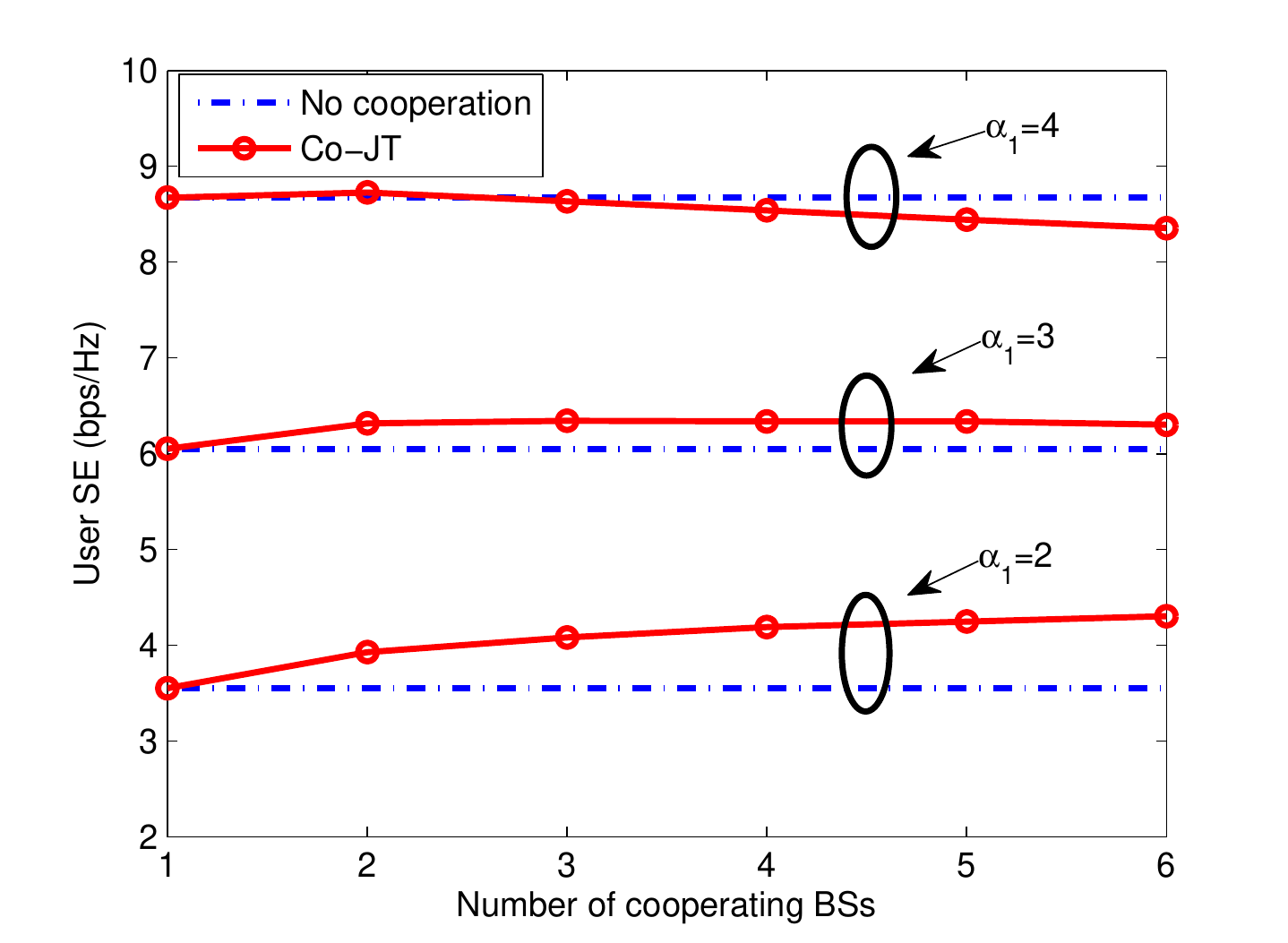}
\caption{User performance with different near-field exponents $\alpha_1$ when $\alpha_2=4$, $\lambda_b=4000/\text{km}^2$, $\lambda_u=200/\text{km}^2$, $R_c$ = 70m.}
\label{fig:PLexponent}
\end{figure}

Fig. \ref{fig:PLexponent} depicts the user SE in different dual-slope path loss environments. We choose a large enough $R_c$ so that all the cooperating BSs are in the near-field. The cooperation gain $G_c$ can be expressed as:
\begin{equation}
\begin{split}
\begin{aligned}
\label{equ:Gain}
G_c&= \frac{\log_{2}\(\frac{|\sum_{j=1}^{N}|h_{j}||d_{j}|^{-\frac{\alpha_1}{2}}|^2}{N|h_{1}|^{2}|d_{1}|^{-\alpha_1}}\)}{\mathcal{R}^o}.
\end{aligned}
\end{split}
\end{equation}

It is easy to prove that the numerator is a decreasing function of $\alpha_1$. Meanwhile, $\mathcal{R}^o$ is an increasing function of both $\alpha_1$ and $\alpha_2$ \cite{Zhang}. Combining the two aspects, a higher path loss exponent returns a lower cooperation gain.

As stated above and shown in the Fig.~\ref{fig:PLexponent}, the case with $\alpha_1=2$ and $\alpha_2=4$ is the most suitable situation for cooperation among the three exemplary cases. In large near-field path loss exponent scenarios where $\mathcal{R}^o$ is already superb, cooperation is not preferable.

\subsection{User and BS Densities}
\label{sec:density}
\begin{figure}[t]
\includegraphics[width=.5\textwidth]{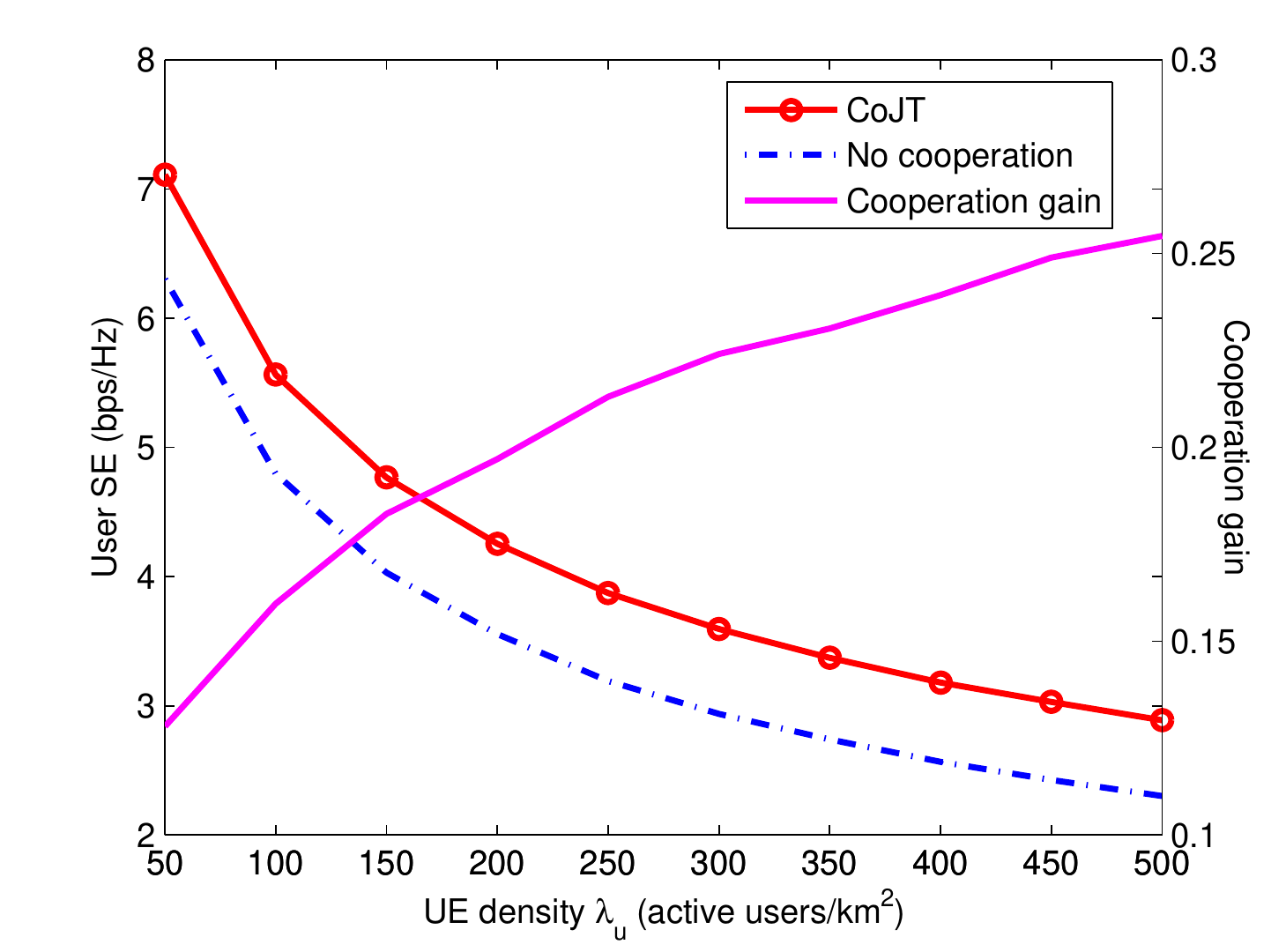}
\caption{Cooperation gain with different user densities when $[\alpha_1,\alpha_2]$ = [2,4], $\lambda_b=4000/\text{km}^2$, $R_c$ = 70m, $N$ = 5.}
\label{fig:UEdensity}
\end{figure}

\begin{figure}[t]
\includegraphics[width=.5\textwidth]{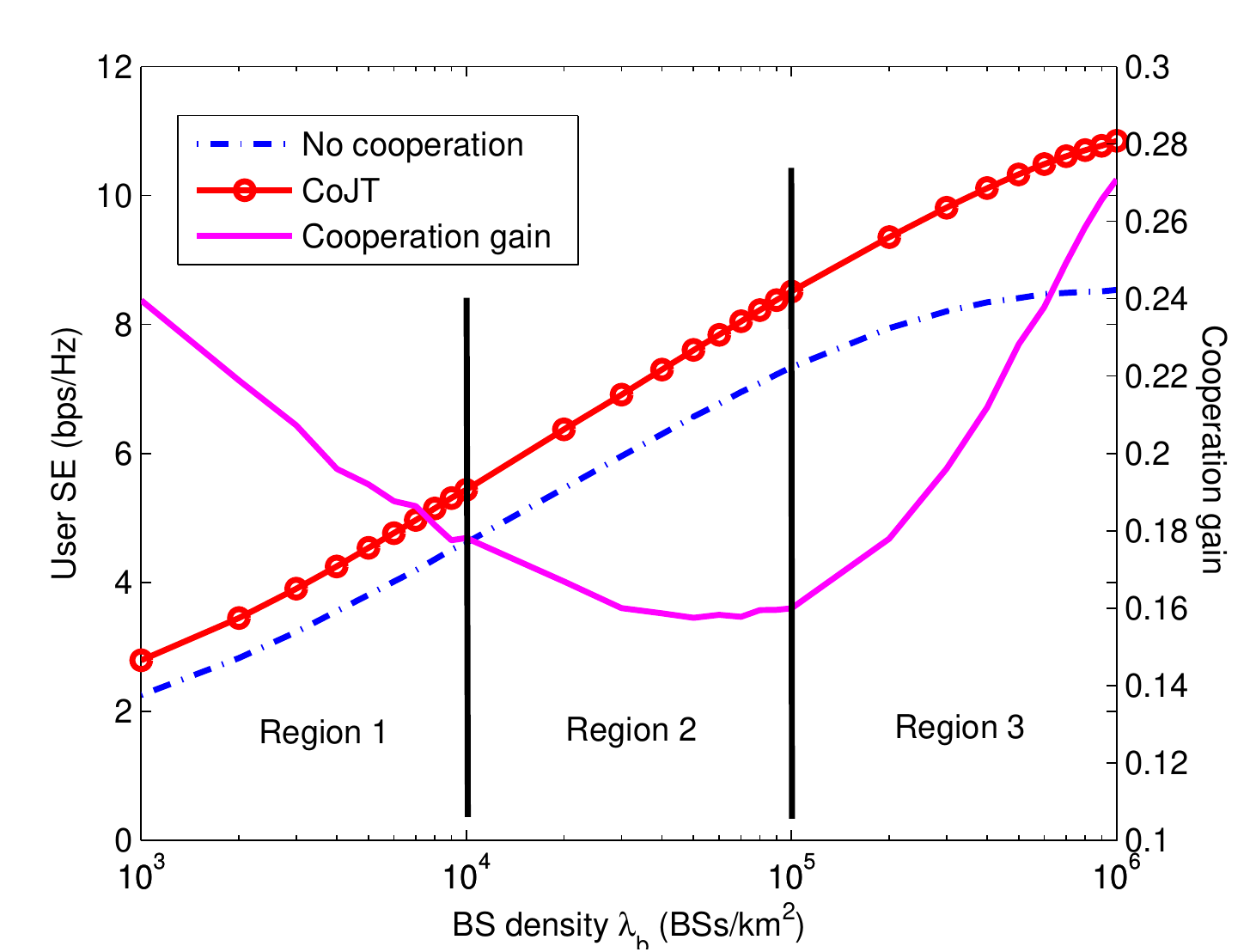}
\caption{Cooperation gain with different BS densities when $[\alpha_1,\alpha_2]$ = [2,4], $\lambda_u=200/\text{km}^2$, $R_c$ = 70m, $N$ = 5.}
\label{fig:BSdensity}
\end{figure}

In this part, we discuss the impact of active user and BS densities on cooperation, shown in Fig. \ref{fig:UEdensity} and Fig. \ref{fig:BSdensity}.

\begin{rem}[Effect of active user density]
As active user density increases, cooperation gain keeps growing while the user SE drops.
\end{rem}

Equation (\ref{equ:Rc2}) can be reused since $\lambda_u$ does not affect $\Delta_{\mathcal{R}}$ same as $R_c$. Besides, $\mathcal{R}^o$ gets smaller as $\lambda_u$ increases due to more interference. Therefore, cooperation performs better with a larger $\lambda_u$. This is similar with the effect of critical distance aforementioned because increasing critical distance is equivalent with increasing user density in the near-field.

\begin{rem}[Effect of BS density]
As BS density increases, cooperation gain is convex-shaped: first decreases logarithmically, then decreases in a lower speed, and finally starts to increase after the BS density reaches a threshold.
\end{rem}

We can write the cooperation gain $G_c$ as:
\begin{equation}
\begin{split}
\begin{aligned}
\label{equ:a}
G_c&= \frac{\log_{2}\(\frac{\mathcal{S}^{\mathit{J}}}{N\mathcal{S}^o}\)}{\mathcal{R}^o} = \frac{\log_{2}\(\frac{|\sum_{j=1}^{N}|h_{j}||d_{j}|^{-\frac{\alpha_1}{2}}|^2}{N|h_{1}|^{2}|d_{1}|^{-\alpha_1}}\)}{\log_{2}\(\frac{|h_{1}|^{2}|d_{1}|^{-\alpha_1}}{I}\)}
\end{aligned}
\end{split}
\end{equation}
where $d_j$ will increase in the order of $\lambda_b^{\frac{1}{2}}$ leading to a fixed numerator. Thus in region 1, the cooperation gain will decrease logarithmically since $\mathcal{R}^o$ will increase logarithmically with $\lambda_b$ \cite{Park2}. In region 2, the decreasing speed of the gain slows down because users start to enter the bounded region where  $\mathcal{S}^o$ and $\mathcal{R}^o$ no longer increase with $\lambda_b$. Finally in region 3, the probability of $d_1 < R_b$ is quite high and $\mathcal{R}^o$ tends to be a constant. Therefore, the cooperation gain increases via further densification.

\begin{figure}[t]
\includegraphics[width=.5\textwidth]{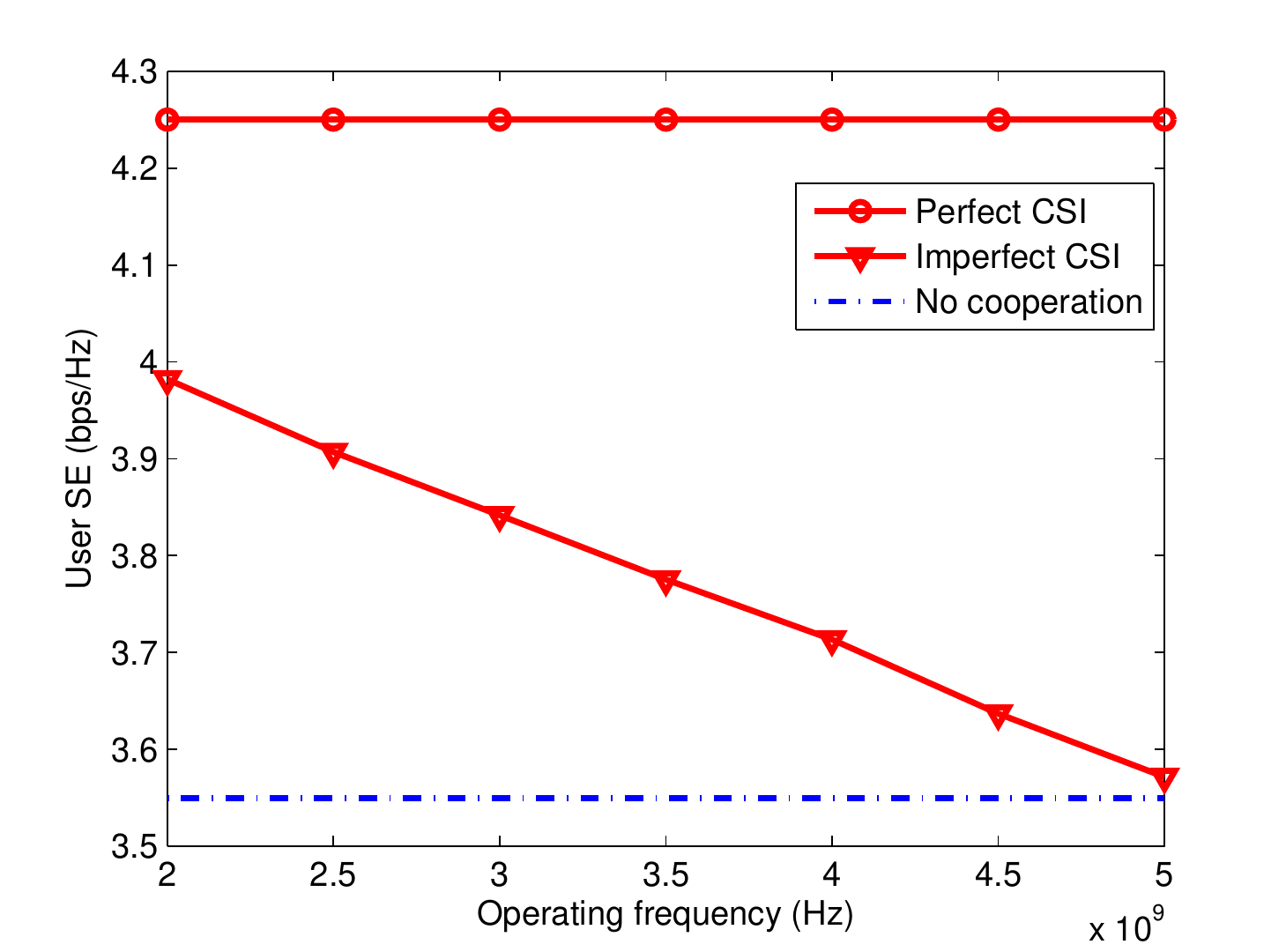}
\caption{Effect of imperfect CSI when $[\alpha_1,\alpha_2]$ = [2,4], $\lambda_b=4000/\text{km}^2$, $\lambda_u=200/\text{km}^2$, $R_c$ = 70m, $N$ = 5.}
\label{fig:imperfectCSI}
\end{figure}

\begin{figure}[t]
\includegraphics[width=.5\textwidth]{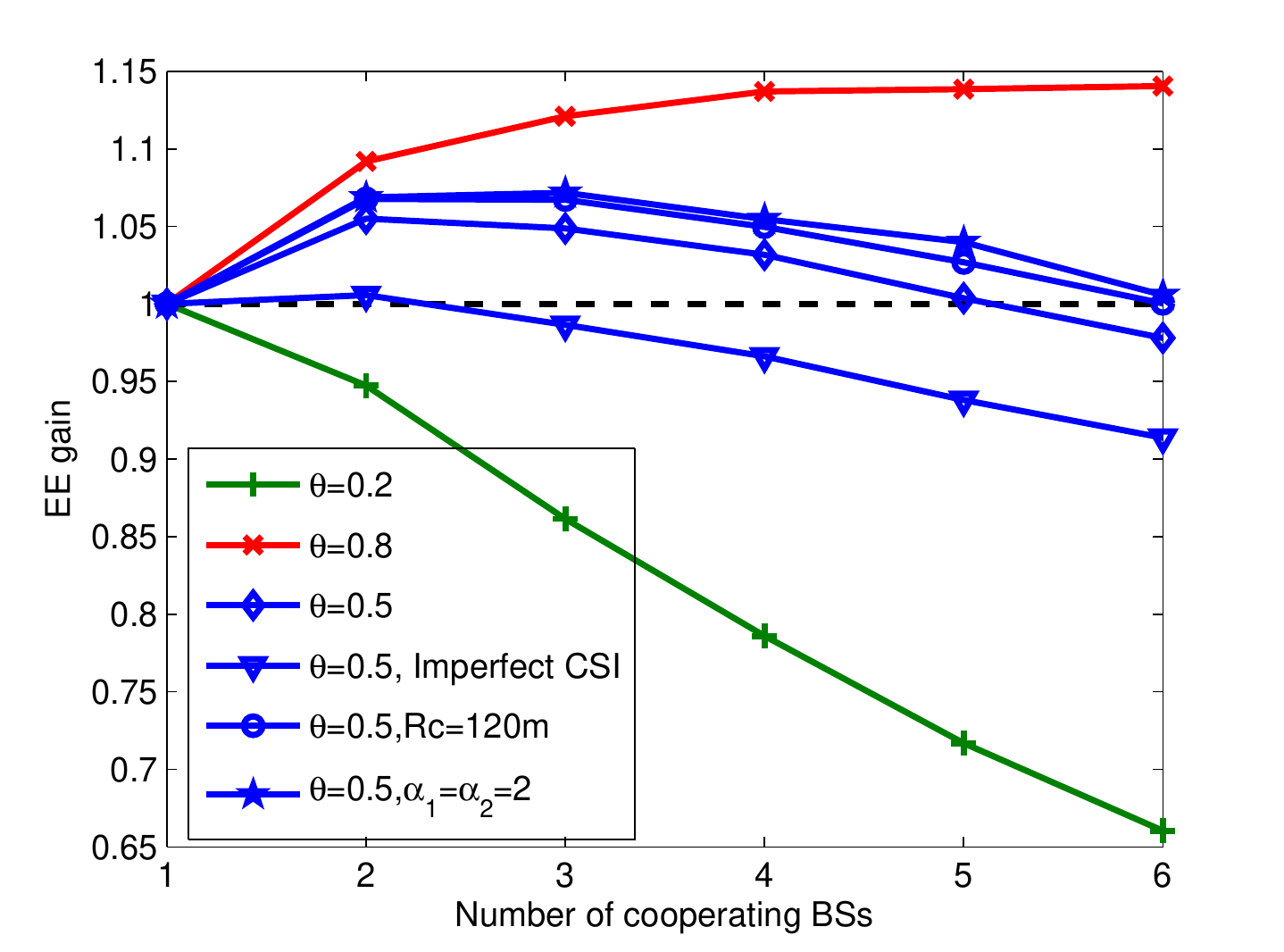}
\caption{Network energy efficiency gain when $[\alpha_1,\alpha_2]$ = [2,4], $\lambda_b=4000/\text{km}^2$, $\lambda_u=200/\text{km}^2$, $R_c$ = 70m if not specified in the figure.}
\label{fig:EE}
\end{figure}

\subsection{Operating Frequency under Imperfect CSI}
We set $T_s$ = 10ms, $v$ = 3km/h and control the operating frequency to identify the effect of imperfect CSI. From Fig.~\ref{fig:imperfectCSI}, cooperation is sensitive to frequency bands. Operating in higher frequency bands leads to a strong Doppler effect and inaccurate channel feedback for moving users. Therefore, the precoders mismatch the instantaneous channel and cannot provide the full cooperation gain.

\subsection{Network Energy Efficiency}
The network EE assessment is present in Fig. \ref{fig:EE}. There exists a tradeoff between SE and EE by waking up the dormant BSs in UDN. Depending on the power consumption model and environmental parameters, EE can even improve when increasing the cooperation number.

When $\theta$ is large, cooperation can improve EE because turning on sleeping BSs will enhance the user SE with little extra energy consumption. On the contrary, a small $\theta$ will cost much more energy consumption resulting in a declining EE. Furthermore, we evaluate EE under favorable environments of cooperation: larger critical distance and small path loss exponents. Both of them can improve EE within certain cooperation number range when $\theta$=0.5. When imperfect CSI is considered, we can hardly achieve an EE improvement due to the SE performance.

\section{Conclusion and Future Work}
\label{sec:Con}
In this paper, we have studied cooperative transmissions in a UDN. Two cooperation schemes, non-coherent JT and coherent JT, are evaluated under a realistic path loss model. We conclude that cooperation is not beneficial without CSI in a UDN. Regarding coherent JT, an environment with longer critical distance and smaller near-field path loss exponent is favorable. Moreover, employing cooperation when user density is higher will be more effective. On the contrary, lower BS density is more helpful except when BS density becomes extremely high, thereby triggering the effect of bounded region. Meanwhile, cooperation is favorable to static users and lower frequency bands operation which allows accurate CSI feedback. Finally, EE can also be improved by cooperation within limited cooperation numbers as long as the extra power consumption is small. Future work can extend to investigation on more advanced BSs with multi-antenna and/or beamforming and seeking the optimal cooperation number.

%Our findings can provide an operation guideline for cooperative transmissions in a UDN. 


%\section*{Acknowledgment}
%Part of this work has been supported by the H2020 project METIS-II co-funded by the EU. The views expressed are those of the authors and do not necessarily represent the project. The consortium is not liable for any use that may be made of any of the information contained therein.




%\bibliographystyle{IEEEtran}

\section*{Acknowledgment}
This research has been partly supported by the H2020 project METIS-II co-funded by the EU. The views expressed are those of the authors and do not necessarily represent the project. It was also supported by Basic Science Research Program through the National Research Foundation of Korea (NRF) funded by the Ministry of Education, Science and Technology (NRF-2015K2A3A1000189).

\ifCLASSOPTIONcaptionsoff
  \newpage
\fi

\bibliographystyle{IEEEtran}
\bibliography{IEEEabrv,bibl}
\end{document}